\documentclass[a4paper, oneside, twocolumn, notitlepage, 10pt]{extarticle_ecoc2015}
\usepackage{ecoc2015}

\newcommand{\nc}{n_{\mathsf{c}}}

\newcommand{\kc}{k_{\mathsf{c}}}

\newcommand{\rr}{\boldsymbol{R}}

\newcommand{\cc}{\boldsymbol{C}}

\newcommand{\dmin}{\mathsf{d}_\mathsf{min}}

\newcommand{\lalone}{\boldsymbol{L}}

\newcommand{\w}{w}
\newcommand{\BB}{\mathsf{B}}

\newcommand{\Eb}{E_\mathsf{b}}
\newcommand{\No}{N_\mathsf{0}}

\def\forcemath#1{\ifmmode #1 \else $#1$\fi}

\begin{document}
	
\selectlanguage{american}    


\title{Iterative Bounded Distance Decoding of Product Codes\\ with Scaled Reliability}


\author{
	Alireza Sheikh\textsuperscript{(1)}, 
	Alexandre Graell i Amat\textsuperscript{(1)}, and
	Gianluigi Liva\textsuperscript{(2)}
}

\maketitle                  

\begin{strip}
	\begin{author_descr}
		
			\textsuperscript{(1)} Dept. of Electrical Engineering, Chalmers University of Technology, Sweden, \uline{asheikh@chalmers.se}
			
			\textsuperscript{(2)} Institute of Communications and
			Navigation, German Aerospace Center (DLR), Germany
		
	\end{author_descr}
\end{strip}

\setstretch{1.071}


\begin{strip}
  \begin{ecoc_abstract}
  We propose a modified iterative bounded distance decoding of product codes. The proposed algorithm is based on exchanging hard messages iteratively and exploiting channel reliabilities to make hard decisions at each iteration. Performance improvements up to $0.26$ dB are achieved.
  
  

\end{ecoc_abstract}
\end{strip}


\section{Introduction}

Product-like codes are a family of codes extensively used in fiber-optic communications, for example in the optical submarine standard ITU-T G.975.1. Product codes (PCs) are a specific class of these codes defined as the set of all rectangular arrays such that each row and each column of the array is a codeword of a given component code. PCs are used as stand-alone codes \cite{Li2011,Justesenetal} or in a concatenation structure (c.f., ITU-T G.975.1, appendix I.5.1) in optical communication systems. As compared to low-density parity-check (LDPC) codes with soft decision decoding,  PCs decoded using iterative bounded distance decoding (IBDD), \cite{Justesenetal} i.e., by iterating BDD decoding of the component codes, yield much lower decoding complexity and power consumption, yet achieving large net coding gains. PCs can be combined with higher order modulations in a so-called  coded modulation (CM) scheme, to boost the spectral efficiency\cite{SheikhHD}. 

In this paper, we propose a novel decoding algorithm for PCs. The proposed algorithm is based on BDD of the component codes and the iterative exchange of binary (i.e., hard) messages between the component decoders. However, compared to conventional IBDD, the proposed algorithm exploits the channel reliabilities to perform the final hard decision at the output of the component decoders. The proposed algorithm can be seen as a modification of the conventional IBDD and we refer to it as IBDD with scaled reliability (IBDD-SR). 
 We show that the proposed decoding algorithm yields superior performance compared to IBDD with no impact in the decoder data-flow, albeit at the expense of a slight increase in complexity. Gains up to $0.26$ dB are observed for a CM system. 


\section{Preliminaries}
We consider binary PCs with Bose-Chaudhuri-Hocquenghem (BCH) codes as component codes. Let $\mathcal{C}$ be an $(\nc,\kc)$ shortened BCH code obtained from a BCH code constructed over the Galois field $\text{GF}(2^\nu)$ with minimum Hamming distance $\dmin$, error correction capability $t \buildrel \Delta \over =  \left\lfloor {\frac{{{\dmin} - 1}}{2}} \right\rfloor$. The (even) block length $\nc$, and information block length $\kc$ are given by $\nc=2^\nu-1-s$ and $\kc=2^\nu-\nu t-1-s$, where $s$ is the shortening parameter. The shortened BCH code is defined by the triplet $(\nu,t,s)$. A (two-dimensional) PC with parameters $(\nc^2,\kc^2)$
can be defined as the set of all $\nc\times \nc$ matrices $\cc=[c_{i,j}]$ such that each row and each column of the matrix is a codeword of the $(\nc,\kc)$ component code. 

We consider bit interleaved coded modulation (BICM) using quadrature amplitude modulation (QAM) over the additive white Gaussian noise (AWGN) channel. Accordingly, $\cc$ is randomly interleaved and the rows of the resulting matrix are concatenated, forming a vector ${\boldsymbol{c'}}=({c'}_1,\cdots,{c'}_{\nc^2})$. The binary reflected Gray mapping is used to map blocks of $m  = \log_2(M)$ bits of ${\boldsymbol{c'}}$ to an $M$-QAM symbol. At the output of the bit-to-symbol mapper, a sequence of symbols $\bm x=(x_1,x_2,\cdots,x_p)$ of length $p  = \nc^2/m$ is obtained. The channel observation at the output of the channel corresponding to the $i$th QAM symbol, is given by $y_{i}=x_{i}+n_{i}$,
where $N_{i}\sim \mathcal{CN}(0,2\sigma^2)$. Given $y_{i}$,  the log-likelihood ratio (LLR) corresponding to the $l$th bit of the corresponding binary label of $x_{i}$ is obtained as   
$L_i^l = \ln \left({\sum\limits_{a \in \mathcal{S}_l^1} {{e^{ - \frac{{|{y_i} - a{|^2}}}{{2{\sigma ^2}}}}}} }/{{\sum\limits_{a \in \mathcal{S}_l^0} {{e^{ - \frac{{|{y_i} - a{|^2}}}{{2{\sigma ^2}}}}}} }}\right)$, where $\mathcal{S}_l^1$ and $\mathcal{S}_l^0$ are the sets of $M$-QAM symbols with a $1$ and $0$ as the $l$th bit of the corresponding labeling, respectively. The sequence of LLRs $L_1^1,\cdots,L_1^m,\cdots,L_p^1,\cdots,L_p^m$ is then reshuffled and de-interleaved forming the $\nc\times \nc$ matrix $\lalone=[L_{i,j}]$, where $L_{i,j}$ is the LLR corresponding to code bit $c_{i,j}$. Let $\rr=[r_{i,j}]$ be the matrix of hard decisions at the output of the channel, i.e., $r_{i,j}$ is obtained taking the sign of $L_{i,j}$ and mapping $- 1 \mapsto 0$ and $+ 1 \mapsto 1$. We denote by $\BB(\cdot)$ the mapping $L_{i,j}\rightarrow r_{i,j}$, i.e., $r_{i,j}=\BB(L_{i,j})$.

\section{Bounded Distance Decoding}\label{BDD}

BDD is a hard decision decoding (HDD) algorithm for linear block codes that corrects all error patterns with Hamming weight up to the error correcting capability of the code, $t$. BDD can be implemented efficiently for BCH codes with small $t$. BDD may introduce decoding miscorrections if the error pattern has Hamming weight larger than $t$ and there exists a codeword with Hamming distance from the received hard detected word less than or equal to $t$. If such a codeword does not exist, a failure is declared. 
We refer to iterative decoding of PCs based on BDD of the component codes as IBDD, which is accomplished by performing BDD of the row and column codes in an iterative fashion.

\section{Iterative Bounded Distance Decoding of PCs With Scaled Reliability}\label{BDD-SRsec}

\begin{figure}[t]
	
	\begin{center}	
		\scalebox{0.75}{		
			
			\begin{tikzpicture}[>=latex']
			\tikzset{Source/.style={rectangle, draw, thick, minimum width=1.2cm, minimum height=0.9cm, rounded corners=2mm}}
			\tikzset{Sourceup/.style={rectangle, draw, thick, minimum width=1.5cm, minimum height=0.9cm, rounded corners=2mm}}		
			\tikzset{Sourcerealimag/.style={rectangle, draw, thick, minimum width=1cm, minimum height=0.3cm, rounded corners=2mm}}		
			\tikzset{Destination/.style={rectangle, draw, thick, minimum width=1.6cm, minimum height=1cm, rounded corners=2mm}}
			\tikzset{Noise/.style={circle, draw, thick, scale=0.5, minimum size=0.1mm}}
			\tikzset{descr/.style={fill=white}}
			\tikzset{Source1b/.style={rectangle, draw=black, thick, minimum width=0.5cm, minimum height=1.2cm, rounded corners=0.5mm}}
			\tikzset{Source2/.style={rectangle, draw, thick, minimum width=0.8cm, minimum height=1.2cm, rounded corners=0.5mm}}
			\tikzset{Sourcebox1/.style={rectangle, draw=cyan, dashed, thick, minimum width=5cm, minimum height=2.6cm,  rounded corners=1mm}}
			\tikzset{Sourcebox2/.style={rectangle, draw=red, dashed, thick, minimum width=2.3cm, minimum height=1.5cm,  rounded corners=1mm}}
			\tikzset{Sourcebox3/.style={rectangle, draw=olive, dashed, thick, minimum width=3cm, minimum height=2cm,  rounded corners=1mm}}		
			
			\node[Source] (GMD) at (1.5,3.5) {\text{BDD}};
			\node[Noise] (Cw) at (1.5,1.5) {\huge $\times$};
			\node[Noise] (Li) at (0,0) {\huge $+$};
			\node[] (signtcn) at (-2.5,3){$\psi_{i,j}^{\mathsf{r},(\ell)}$};	
			\node[Source,rotate=45] (sign) at (-1.2,1.2) {$\BB(\cdot)$};
			\node[] (GMDnext) at (-4,3.5) {};
			\node[] (GMDnext1) at (-4,3.9) {\text{$j$\text{th column}}};			
			\node (checknode1) at (-3.9,3.6) {\text{decoder}};				
			\node[Sourcebox3,fill=olive, fill opacity=0.35] (VN) at (-3.5,3.4) {};				
			\node (variablenode) at (3.25,0.72) {$(c_{i,j})$};
			
			\draw [->] (GMD) --  node [right] {$\mu^{\mathsf r}_{i,j} \in \{\pm 1,0\}$} (Cw);
			\draw [->] (Cw) --  (Li);
			\draw [->] (Li) --  (sign);
			\draw [->] (sign) --  (GMDnext);		
			\draw [->] (0,-1.5) --  node [right] {$L_{i,j}$} (Li);
			\draw [->] (3.5,1.5) --  node [above] {$w$} (Cw);
			\node (checknode) at (2.3,-0.4) {$i$\text{th row}};
			\node (checknode) at (2.3,-0.7) {\text{decoder}};				
            \draw[Periwinkle,dashed,fill=Periwinkle,fill opacity=0.3](4,4.2) --  (0.6,4.2) -- (-2.2,2) -- (-2.2,-1.7) -- (4,-1.7) -- (4,4.2);
            			

			\draw [->] (1.5,4.4) -- (GMD);      
			\node[] (inpuut) at (1.5,4.6) {$\bm{\psi}_{i,:}^{\mathsf{c},(\ell-1)}$};          
			
			\end{tikzpicture}
		}
	\end{center}
	\vspace{-1ex} 
	\caption{Block diagram of IBDD-SR for the hard decision on bit $c_{i,j}$ by the $i$th row decoder at iteration $\ell$. $\bm{\psi}_{i,:}^{\mathsf{c},(\ell-1)}$ is the $i$th row of $\bm{\Psi}^{\mathsf{c},(\ell-1)}$.}
	\label{messagepasing1}
	\vspace{-0.5ex}
\end{figure}
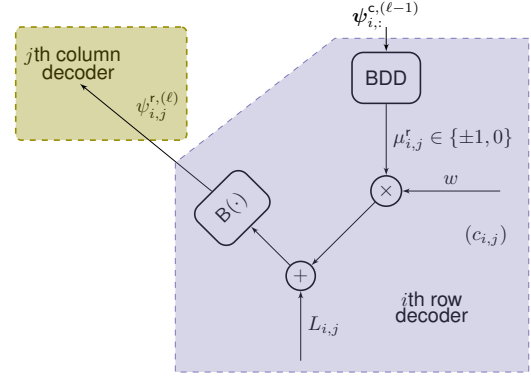

The proposed IBDD-SR works as follows. Without loss of generality, assume that decoding starts with the decoding of the row codes and, in particular, consider the decoding of the $i$th row code at iteration $\ell$. 
Let $\bm{\Psi}^{\mathsf{c},(\ell-1)}=[\psi_{i,j}^{\mathsf{c},(\ell-1)}]$ be the matrix of hard decisions on code bits $c_{i,j}$ after the decoding of the $\nc$ column codes at iteration $\ell-1$. Row decoding is then performed based on $\bm{\Psi}^{\mathsf{c},(\ell-1)}$. We assume that the output of the BDD stage of the $i$th row component code  corresponding to code bit $c_{i,j}$, denoted by $\mu_{i,j}^{\mathsf r}$, takes values on a ternary alphabet, $\mu_{i,j}^{\mathsf r} \in \{\pm1, 0 \}$, where $-1$ corresponds to the hard-decoded bit $0$, $+1$ corresponds to the hard-decoded bit $1$, and $0$ corresponds to  unsuccessful BDD of the $i$th component row code. 
Then, the hard decision on $c_{i,j}$ made by the $i$th row decoder is formed as
\begin{equation*}\label{eq:BDDchrel_VN_scale}
\scriptsize{	
\psi_{i,j}^{\mathsf{r},(\ell)}=
\BB(\w \cdot \mu_{i,j}^{\mathsf r} + L_{i,j}) }, 
\end{equation*}
where ties can be broken with any policy. Parameter $w$ is a scaling parameter that needs to be properly optimized. After decoding of the $\nc$ row codes at decoding iteration $\ell$, the matrix $\bm{\Psi}^{\mathsf{r},(\ell)}=[\psi_{i,j}^{\mathsf{r},(\ell)}]$ is passed to the $\nc$ column decoders, and column decoding based on $\bm{\Psi}^{\mathsf{r},(\ell)}$ is performed. As before, we assume that the output of the BDD stage of the $j$th column component code corresponding to code bit $c_{i,j}$, denoted by $\mu_{i,j}^{\mathsf c}$, takes values on $\{\pm1, 0 \}$, where $+1$, $-1$ and $0$ have the same meaning as before but for the $j$th component column decoder. 
Then, the hard  decision on $c_{i,j}$ made by the $j$th column decoder at iteration $\ell$ is formed as
\begin{equation*}\label{eq:BDDchrel_VN_scale}
\scriptsize{
		\psi_{i,j}^{\mathsf{c},(\ell)}=\BB(\w \cdot \mu_{i,j}^{\mathsf c} + L_{i,j})}.
\end{equation*}
After decoding of the $\nc$ column codes at decoding iteration $\ell$, the matrix $\bm{\Psi}^{\mathsf{c},(\ell)}=[\psi_{i,j}^{\mathsf{c},(\ell)}]$ is passed to the $\nc$ row decoders for the next decoding iteration. The schematic diagram of the proposed IBDD-SR  is shown in Fig.~\ref{messagepasing1}.
The crucial modification in IBDD-SR with respect to conventional IBDD is that the hard decisions passed between component decoders are not simply the result of the BDD of the component codes, but are made on the sum of a scaled version of the output of the BDD decoder and the channel LLR.  
Therefore, the channel reliabilities are exploited to make the final hard decisions at each row and column decoding stage. 
Intuitively, since the channel reliability is exploited in the hard decisions at each row and column decoding, in the case that the reliability of the channel is high ($|L_{i,j}|$ is large) and conventional IBDD introduces miscorrections, the modified algorithm may combat the possible miscorrections. We highlight that the proposed algorithm exchanges binary (hard) messages, hence the decoder data-flow is the same as that of conventional IBDD. This comes at a minor increase in complexity (a product and a sum for each code bit and iteration) and some memory increase as the channel LLRs need to be stored.

\section{Simulation Results}
\begin{figure}[t] \centering 
	\includegraphics[width=\columnwidth]{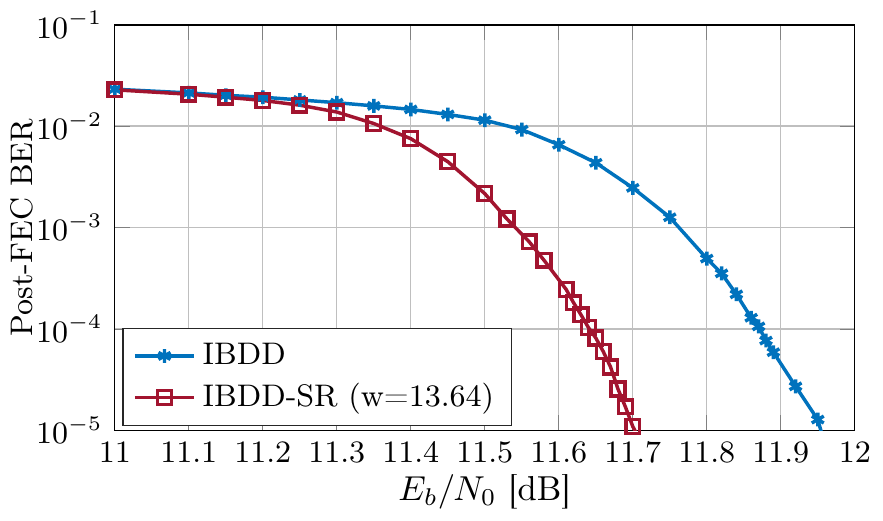}  
	\vspace{-3ex}
	\caption{BER for $\mathcal{C}_{2}$ with $64$-QAM.}\vspace{-0.5ex}
	\label{v8t3s63_product_64QAM} 
	\hspace{0.05mm}	
	\includegraphics[width=\columnwidth]{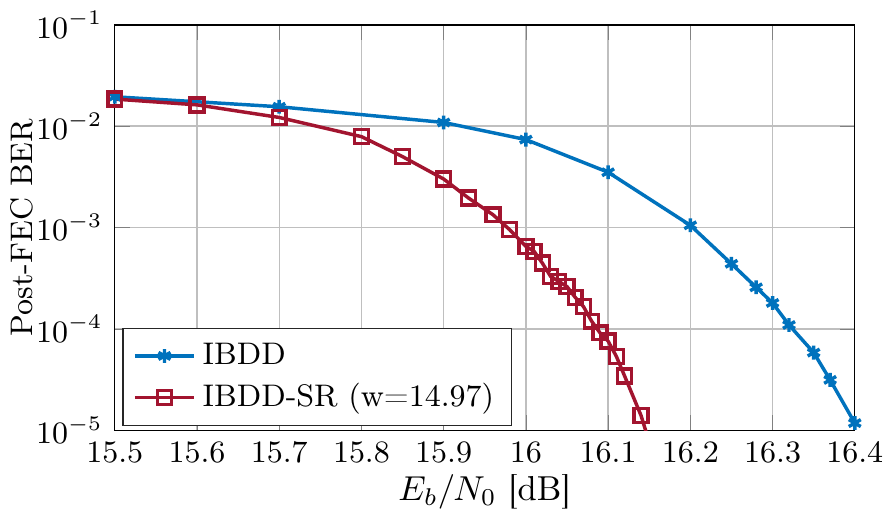}  
	\vspace{-3ex}
	\caption{BER for $\mathcal{C}_{2}$ with $256$-QAM.} \vspace{-0.5ex}
	\label{v8t3s63_product_256QAM}		
\end{figure}

In order to assess the performance of IBDD-SR, it is required to optimize the scaling parameter $w$. For a given component code $(\nu,t,s)$, we simulate the performance of the PC for $w=1$ and find the corresponding threshold, $(\Eb/\No)^*$ (with $\Eb$ being the energy per information bit and $\No$ the single-sided noise spectral density), where the bit error rate (BER) is  $10^{-4}$. We search numerically for the value of $w$ that minimizes $(\Eb/\No)^*$ and use it for performance evaluation in the entire range of $\Eb/\No$. We remark that in our simulations $w$ is fixed across iterations, since by allowing the variation of $w$ over iterations we found a negligible BER improvement.
\begin{figure}[t] \centering 
	\includegraphics[width=\columnwidth]{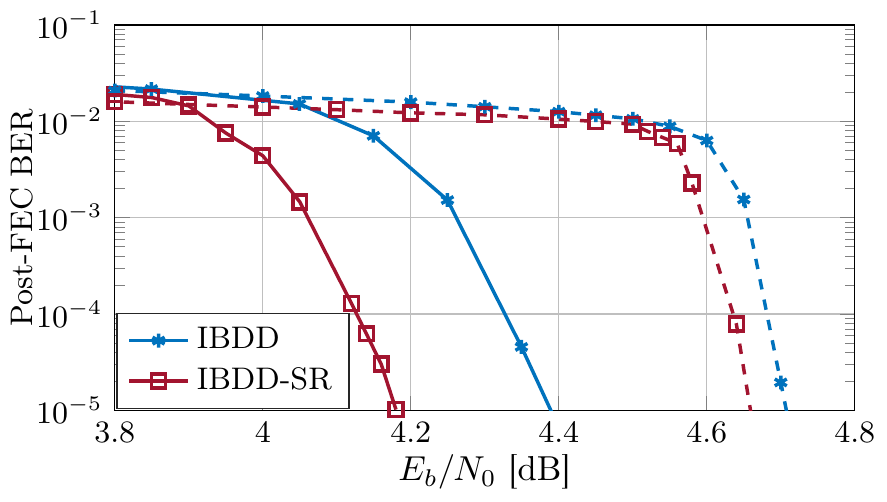}  
	\vspace{-3ex}
	\caption{BER for $\mathcal{C}_{1}$ (dashed, $w=1.44$), and $\mathcal{C}_{2}$ (solid, $w=1.13$) with BPSK.} \vspace{-0.5ex}
	\label{compare_BPSK}
	\hspace{0.05mm}
	\includegraphics[width=\columnwidth]{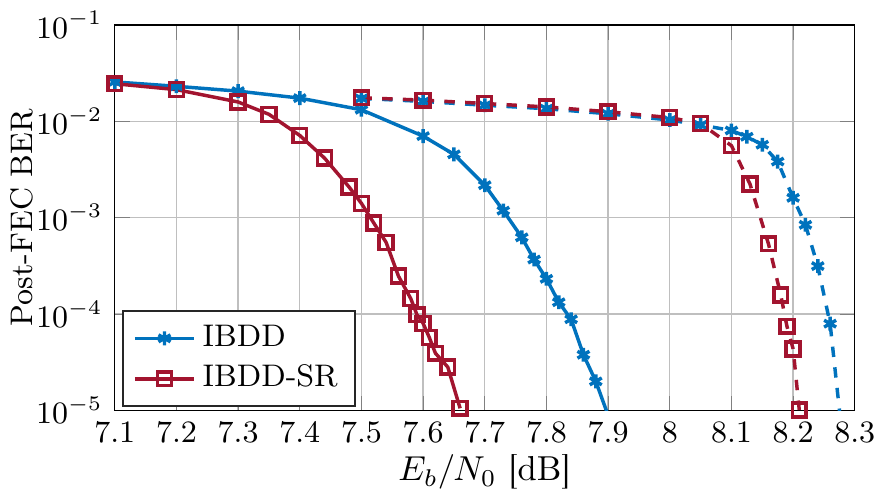}  
	\vspace{-3ex}
	\caption{BER for $\mathcal{C}_{1}$ (dashed, $w=12.82$), and $\mathcal{C}_{2}$ (solid, $w=8.88$) with $16$-QAM.} \vspace{-1ex}
	\label{compare_16QAM}		 
\end{figure}
We consider two PCs, $\mathcal{C}_{1}$ and $\mathcal{C}_{2}$, with component codes $(9,4,7)$\cite{Hager15b} and $(8,3,63)$\cite{Hager15b}, respectively, and corresponding code rates $0.8622$ and $0.7656$ and $t=4$ and $t=3$.
In Fig.~\ref{compare_BPSK} and Fig.~\ref{compare_16QAM}, the performance of CM for PCs $\mathcal{C}_{1}$ and $\mathcal{C}_{2}$ decoded using IBDD-SR and IBDD with $10$ iterations for binary phase shift keying (BPSK) and $16$-QAM is depicted. As can be seen, the performance gains of IBDD-SR compared to IBDD at BER of $10^{-5}$ are $0.05$ dB and $0.21$ dB for BPSK with $\mathcal{C}_{1}$ and $\mathcal{C}_{2}$, respectively, while the gains are $0.065$ dB and $0.23$ dB for $16$-QAM. For $t=3$ IBDD-SR yields a significant gain, while for $t=4$ the gain is very small. The intuition is that for larger $t$, the probability of miscorrection of the component code is reduced, hence, the performance gain of IBDD-SR compared to IBDD is reduced. In Fig.~\ref{v8t3s63_product_64QAM} and Fig.~\ref{v8t3s63_product_256QAM}, we plot the performance of $\mathcal{C}_{2}$ decoded with $10$ iterations for $64$-QAM and $256$-QAM, respectively. The  performance gains of IBDD-SR compared to IBDD at a BER of $10^{-5}$ are $0.25$ dB and $0.26$ dB for $64$-QAM and $256$-QAM, respectively. The performance improvement of IBDD-SR over iterative IBDD is slightly higher than for BPSK. 
Our simulations show similar gains ($\sim 0.26$ dB) for component codes with $t<3$. Note that low values of $t$ (e.g., $t=3$) are interesting for optical communications, since a smaller $t$ entails a lower decoding complexity (and higher throughput).

\section{Conclusion}

A novel decoding algorithm for product codes, based on BDD of the component codes, is proposed. As conventional IBDD, the proposed algorithm exchanges binary (hard) messages between component decoders. The key novelty with respect to conventional IBDD is that the hard decisions on the code bits at each decoding stage exploit the channel LLRs. The performance gains of the novel IBDD-SR over conventional IBDD are considerable for component codes with $t\le3$. Gains up to $0.26$ dB are observed compared to IBDD for $256$-QAM modulation.  
Currently, we are extending IBDD-SR to staircase codes.

\section{Acknowledgment}
{This work was supported by the Knut and Alice Wallenberg and the Ericsson Research Foundations.}


\end{document}